\documentclass[a4paper]{aa}
\usepackage{txfonts}
\usepackage[usenames,dvipsnames,svgnames,table]{xcolor}
\usepackage{graphicx}
\usepackage{url}

\usepackage{hyperref}
\usepackage[hyphenbreaks]{breakurl}

\outer\def\gtae {$\buildrel {\lower3pt\hbox{$>$}} \over 
{\lower2pt\hbox{$\sim$}} $}
\outer\def\ltae {$\buildrel {\lower3pt\hbox{$<$}} \over 
{\lower2pt\hbox{$\sim$}} $}
\newcommand{\Msun} {$\mathrm{M}_{\odot}$}
\newcommand{\Rsun} {$\mathrm{R}_{\odot}$}

\newcommand{\src}{XMMJ0510-67}

\begin{document}

\title{Detection of a 23.6 min periodic modulation in the optical counterpart of 3XMMJ051034.6--670353}
\author{G. Ramsay\inst{1}, T.\ R. Marsh\inst{2}, T. Kupfer\inst{3}, V. S. Dhillon\inst{4,5},
D. Steeghs\inst{2},  P. Woudt\inst{6}, P. Groot\inst{7}}
\authorrunning{Ramsay et al.}
\titlerunning{Periodic optical modulation found in 3XMMJ051034.6--670353}
\institute{Armagh Observatory and Planetarium, College Hill, Armagh, BT61 9DG,
  UK\label{inst1} \and
  Department of Physics, University of Warwick, Coventry CV4 7AL, UK\label{int2}\and  
  Division of Physics, Mathematics and Astronomy, California Institute of Technology, Pasadena, CA 91125, USA\label{inst3}\and
  Department of Physics \& Astronomy, University of Sheffield, Sheffield, S3 7RH, UK\label{int4}\and
  Instituto de Astrofísica de Canarias, E-38205 La Laguna, Tenerife, Spain\label{int5}\and
 Inter-University Institute for Data Intensive Astronomy, Department of Astronomy, University of Cape Town, Private Bag X3, Rondebosch 7701, South Africa\label{int6}\and
Department of Astrophysics/IMAPP, Radboud University, PO Box 9010, NL-6500 GL Nijmegen, Netherlands\label{int7}\\
  \email{gavin.ramsay@armagh.ac.uk}}


\abstract{We present high speed optical photometric observations made
  using the NTT and ULTRACAM of the optical counterpart of
  3XMMJ051034.6--670353, which was recently identified as an X-ray
  source showing a modulation on a period of 23.6 min. Although the
  optical counterpart is faint ($g$=21.4), we find that the
  $u^{'}g^{'}r^{'}$ light curves show a periodic modulation on a
  period which is consistent with the X-ray period. We also obtained
  three low resolution spectra of 3XMMJ051034.6--670353 using the
  Gemini South Telescope and GMOS. There is no evidence for strong
  emission lines in the optical spectrum of 3XMMJ051034.6--670353. We
  compare and contrast the optical and X-ray observations of
  3XMMJ051034.6--670353 with the ultra compact binaries HM Cnc and
  V407 Vul. We find we can identify a distribution of binary masses in
  which stable direct impact accretion can occur.}


\maketitle

\section{Introduction}

Twenty years ago, two X-ray sources were discovered during the course
of the {\sl ROSAT} All-Sky Survey which showed a characteristic
`on/off' flux profile which were modulated on a period of 321 sec (RX
J0806+15, \citet{Israel1999}) and 569 sec (RX J1914+24,
\citet{Motch1996}).  The optical counterparts were found to show a
modulation on the same period as the X-ray flux (HM Cnc:
\citet{Ramsay2002}, \citet{Israel2002}); V407 Vul:
\citet{Ramsay2000}).  It was later found that the peak of the X-ray
and optical flux profile in both systems is shifted by $\sim$0.2
cycles (\citet{Barros2007}. Because of their faintness (both
$V\sim$20), and their short period, it took some time to confirm that
the observed 321 sec period seen in HM Cnc was the binary orbital
period \citep{Roelofs2010} and therefore the most compact known binary
star. Today, HM Cnc and V407 Vul are still the two most compact
stellar binary systems known.

HM Cnc and V407 Vul are at the very short period end of the orbital
period distribution of `AM CVn' binaries (5--70 min). They consist of
two degenerate (or semi-degenerate) stars orbiting around a common
center of gravity (see \citet{Solheim2010} for a review). As the less
massive secondary star fills its Roche Lobe, material flows from the
secondary star towards the more massive star. In the case of HM Cnc
and V407 Vul, it is still an open question as to whether the material
from the mass donor forms a small accretion disc or that accretion
occurs through `direct impact' \citep{MarshSteeghs2002} or magnetic
accretion \citep{Cropper1998}, \citep{Barros2007}, although the radial
velocity study of \citep{Roelofs2010} seems to favour the direct
impact scenario in HM Cnc. It is thought that a small fraction of AM
CVn's may result in a merger and potentially a SN 1a or .1a explosion
(see \citet{Kilic2014}). They are predicted to be the brightest known
sources of persistent gravitational waves and are `verification'
sources for any space-based gravitational wave observatory such as
LISA (e.g. \citet{Nelemans2013}, \citet{Korol2017}).

Currently there are 52 known AM CVn binaries, but none have similar
characteristics to HM~Cnc or V407~Vul. For instance, ES~Cet, with an
orbital period of 620 sec \citep{Warner2002}, only a little
longer than V407~Vul's period, has very different photometric
properties and accretes via a disc. This suggests that sources like HM
Cnc and V407 Vul are rare, perhaps because the direct impact phase
lasts a short time (a few million years or less). Given the great
interest in these sources from the gravitational wave community
(amongst others) it is of great interest to find more examples of
these ultra-short period binaries.

A recently discovered X-ray source (3XMMJ051034.6--670353, \src\ for
brevity) in the {\sl XMM-Newton} EXTraS survey
\citep{Haberl2017}. shows an X-ray modulation on a period of 23.6 min
and an on/off X-ray light curve very similar to HM Cnc and V407 Vul
but with a lower flux. The fact that the most likely optical
counterpart is faint ($g$=21.3, Haberl et al. 2017) has made it
difficult to determine if the optical flux is modulated on the same
period as the X-ray period.

In this paper, we present ULTRACAM photometric observations made using
the ESO NTT which show that the optical counterpart of \src\ also
shows period variability on a timescale of 23.6 min (\S
\ref{ultracam}). In \S \ref{gemini} we present an optical spectrum of
\src\ obtained from the Gemini South Telescope. In \S \ref{discuss} we
outline the possible interpretation of these observations and their
implications.

\section{ULTRACAM photometric observations}
\label{ultracam}

Observations of \src\ were made using the ULTRACAM multi-band imager
which is a visitor instrument at ESO's 3.5m New Technology Telescope
(NTT) on La Silla, Chile. ULTRACAM has essentially zero readout time
and allows fast simultaneous three-colour photometric measurements of
faint targets \citep{Dhillon2007}. Observations started on 2017 Nov
22 at 02:42 UT and lasted 4.04 hrs. Conditions were clear and seeing
was $\sim1.0^{''}$. The exposure time for the $g^{'}$ ($\sim$4000-5500
\AA) and $r^{'}$ (5500--6900 \AA) bands were 20 sec each, and 60 sec
in the $u^{'}$ band (3600-3900 \AA).

\begin{figure}
\begin{center}
\setlength{\unitlength}{1cm}
\begin{picture}(7,7)
\put(-2,-3.5){\includegraphics{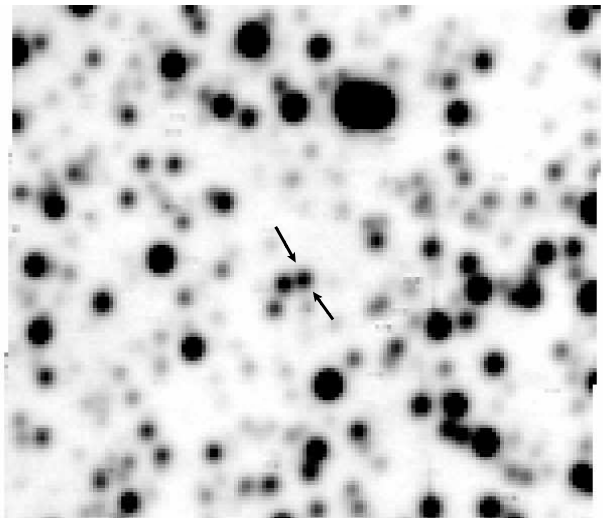}}
\end{picture}
\end{center}
\caption{The co-added $g^{'}$ band image of the field surrounding \src\
  (which is marked by two arrows and is source `1' in Fig 9 of Haberl et
  al. 2017). It was made from all 727 $g^{'}$ band images each with an
  exposure of 20 sec. The image is 1$\times$1 arcmin across with North
  being to the top and East to the left. We estimate the depth of the
  image reaches $g^{'}\sim$23.6.}
\label{finding} 
\end{figure}

In Figure \ref{finding} we show the co-added $g$ band
$1^{'}\times1^{'}$ image centered on \src. Although the field is out
of the Galactic plane ($b=-34.6^{\circ}$), it is quite crowded and a
slightly brighter star ($g^{'}$=21.0) is only 1.8$^{''}$ from
\src\ \citep{Haberl2017}. \citet{Haberl2017} also show that \src\ has a
  colour $g-r$=--0.03$\pm$0.02.  Comparing stars in the field with the
  Skymapper $uv$ band magnitudes \citep{Scalzo2017} we estimate that
  \src\ gives $u-g\sim$0.1--0.2, where the uncertainty stems from the
  inexact relationship between the ULTRACAM $u^{'}$ filter and the
  Skymapper passbands.

\begin{figure*}
\begin{center}
\setlength{\unitlength}{1cm}
\begin{picture}(12,10.5)
\put(16,-0.8){\includegraphics{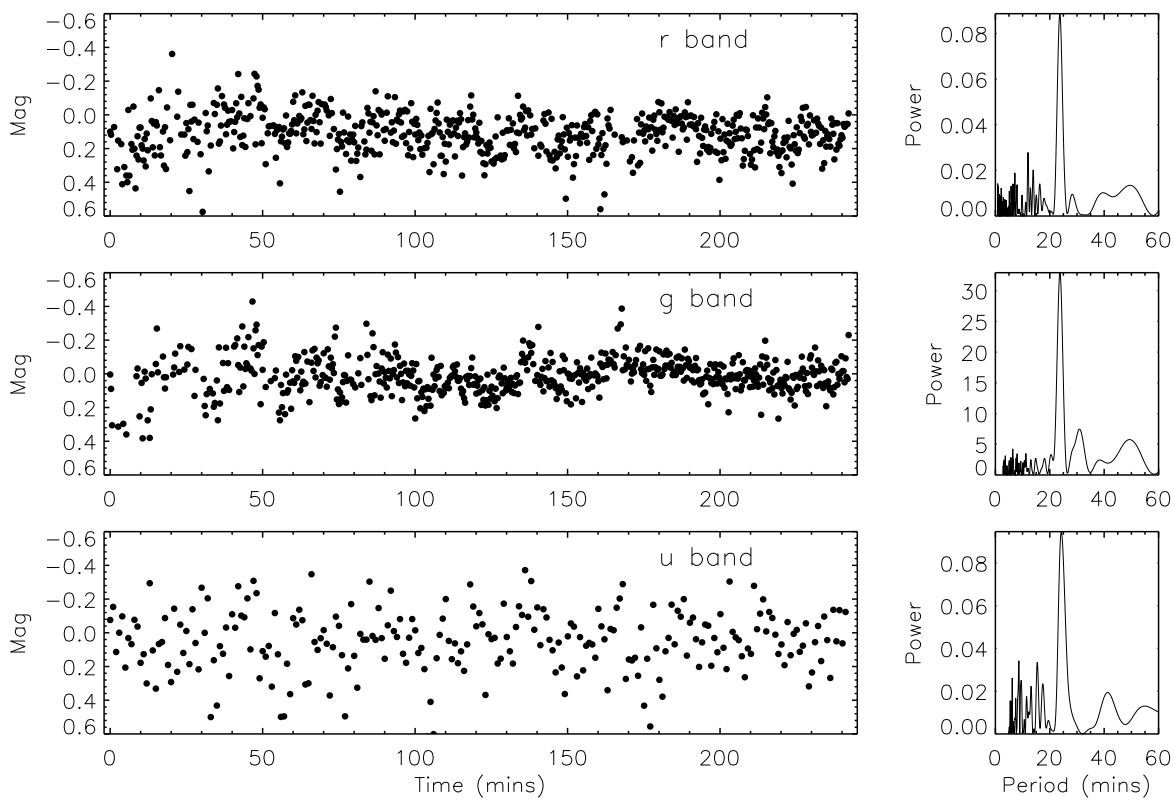}}
\end{picture}
\end{center}
\caption{Left hand panels: The light curve of \src\ made using the NTT
  and ULTRACAM in the $r^{'}$ band (upper), $g^{'}$ band (middle) and
  $u^{'}$ band (lower) and normalised so the mean magnitude is
  0.0. Right hand panels: the Lomb-Scargle periodogram of the light
  curves shown in the left hand panels.}
\label{rbandlight} 
\end{figure*}

For the $g^{'}$ and $r^{'}$ data we used the profile fitting package
{\tt psfex} \citep{Bertin2011}. There were significantly fewer stars
in the $u^{'}$ band image and \src\ was faint and we used standard
fixed aperture photometry using {\tt autophotom} \citep{Eaton2009}.
For the $g^{'}$ and $r^{'}$ data we compared the magnitudes of all
stars in each image with that derived from the coadded frame made from
the five images with the best seeing. We used the brighter stars to
determine an offset between each image and the coadded image to derive
the relative brightness of all stars in the field. We also detrended
all light curves by subtracting a low order polynomial. We obtained
the Lomb Scargle power spectrum for each light curve (both before and
after detrending): \src\ stood out as being variable on a period of
$\sim$23.6 min in both the $g^{'}$ and $r^{'}$ data (both before and
after detrending). We show the $g^{'}$ and $r^{'}$ band light curves
for \src\ in the left hand panels of Figure \ref{rbandlight} and the
Lomb Scargle periodogram in the right hand panels of Figure
\ref{rbandlight}. For the $u^{'}$ band images we obtained a light
curve of \src\ by obtaining differential photometry with nearby
comparison stars. The star 1.8$^{''}$ to the East of \src\ shows no
evidence of periodic variability.

\begin{figure}
\begin{center}
\setlength{\unitlength}{1cm}
\begin{picture}(6,6)
\put(-2,-6.5){\includegraphics{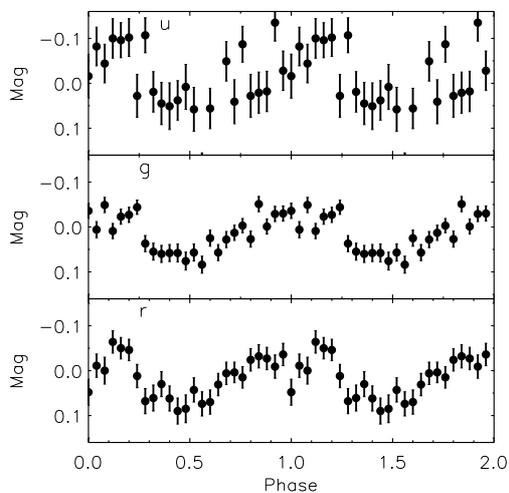}}
\end{picture}
\end{center}
\caption{The light curves of \src\ in the $u^{'}g^{'}r^{'}$ filters
  folded on a period of 23.63 min and $T_{o}= HMJD\ 58079.112$.}
\label{fold} 
\end{figure}

We performed a least-squares fit to the light curves using a sinusoid
to determine the uncertainty of the period (from the covariance
matrix) and we show these in Table \ref{periods}. The periods in all
three optical bands are consistent to within $< 3\sigma$ with the
X-ray period (23.63$\pm$0.02 min, Haberl et al. 2017). We folded the
light curves on a period of 23.6 min and show these in Figure
\ref{fold}. The semi-amplitude of the modulation is $\sim$0.08-0.10
mag at $u^{'}g^{'}r^{'}$ bands: this is lower than HM~Cnc (a
semi-amplitude of $\sim$0.15 mag) and also the $u$ band amplitude
observed in V407 Vul \citep{Barros2007}. In the non-detrended light
curves, there is some evidence for a longer period trend (especially
in the $g^{'}$ band). Further observations are required to determine
if this is intrinsic to the source or perhaps a colour effect.

\begin{table}
\begin{center}
\begin{tabular}{lcr}
\hline
Filter & Period & Amplitude \\
       & (min)  & (mag) \\
\\
\hline
$u^{'}$ & 24.24$\pm{0.28}$ &  0.10 \\
$g^{'}$ & 23.79$\pm{0.16}$ &  0.08 \\
$r^{'}$ & 23.63$\pm{0.17}$ &  0.09 \\
\hline
\end{tabular}
\end{center}
\caption{The period and amplitude of the modulation in the detrended
  $u^{'}g^{'}r^{'}$ light curves of \src. The period derived from the
  $g^{'}r^{'}$ data is entirely consistent with the X-ray period
  (23.63$\pm$0.03 min, \citet{Haberl2017}. The $u^{'}$ band data is of
  lower quality and the uncertainty on the period is less reliable
  than the $g^{'}r^{'}$ estimates.}
\label{periods}
\end{table}

\section{Gemini spectroscopic observations}
\label{gemini}

We obtained low resolution spectroscopic observations of \src\ using
the 8.1m Gemini South Telescope in Cerro Pachon, Chile, using GMOS in
long slit mode. We obtained three exposures each lasting 1200 sec on
the night of 2017 March 26/27. We used the R400 grating plus G5325
filter with a slit of 1.5$^{''}$ put in north-south direction which
gave a resolution of $R\sim$390--470 (blue to red). By the third
exposure the seeing had degraded but the sequence was continued to
allow the conclusion of the observations. The data were reduced using
standard routines and we used the flux standard CD-32 9927 to remove
the instrumental response. As shown in Figure \ref{finding} there is a
star which is slightly brighter than \src\, 1.8$^{''}$ to the
East. Given the width of the slit and the poor seeing, there is a
potential issue that the spectrum has been contaminated by the nearby
star.

\begin{figure}
\begin{center}
\setlength{\unitlength}{1cm}
\begin{picture}(7,6.5)
\put(-1,0){\includegraphics{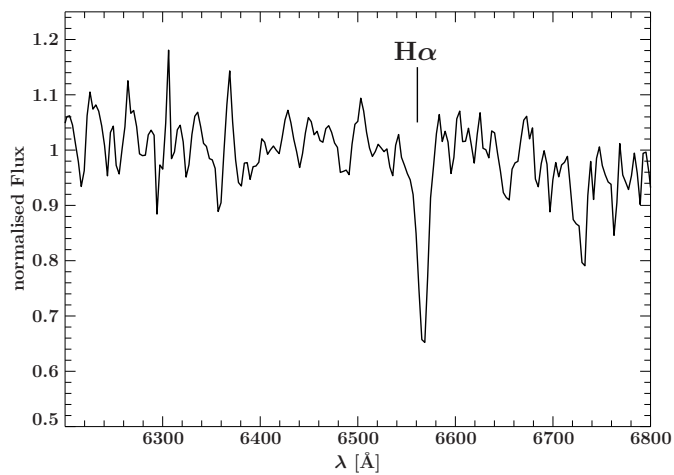}}
\end{picture}
\end{center}
\caption{The co-added optical spectrum of \src\ obtained using the
  Gemini South Telescope. Although the signal-to-noise was low we
  detected an absorption feature at a wavelength consistent with
  H$\alpha$.}
\label{Halpha}
\end{figure}

We extracted a spectrum from each observation and to increase the
signal-to-noise ratio (SNR) we co-added the individual exposures
resulting in an average spectrum with a SNR$\approx10$. There is no
evidence of strong emission lines in the average spectrum. We find
only evidence of a narrow feature at a wavelength consistent with
H$\alpha$ in absorption (see Figure \ref{Halpha}) and measure an
equivalent width of $14\pm2\AA$ of that line. However, we note that a
He{\sc ii} line at 6560$\AA$ has been seen in some AM CVn systems (see
for instance \citet{Green2018}).  A higher signal-to-noise spectrum is
needed to search for the presence of H$\beta$ and H$\gamma$ which
would be expected if the line near 6563$\AA$ is indeed hydrogen.
Although the spectra are of low signal-to-noise, it is clear that
there are no strong emission lines in the optical spectrum of
\src\ and there is a possibility that hydrogen is present.

\section{Discussion}
\label{discuss}

We have found that the optical counterpart of \src\ shows a modulation
on a period 23.6 min) which is consistent with that detected in X-rays
\citep{Haberl2017}. \src\ therefore shows clear similarities with
the two shortest known stellar binary systems HM Cnc and V407 Vul: all
three show a periodic modulation at the same period in optical and
X-ray wavelengths and all show a very soft X-ray spectrum.  In HM Cnc
and V407 Vul there is a phase offset between the X-ray and optical
light curves. Because of the uncertainty in the X-ray period of
\src\ we are unable to phase the X-ray and optical data on the same
ephemeris.

As noted in \S \ref{gemini} the optical spectrum of \src\ shows no
evidence for strong emission lines. This is very different to ES Cet
with a period of 10.3 min which shows strong helium emission lines
\citep{Espaillat2005}. However, CR Boo with a period of 24.5 min shows
very weak absorption features in quiescence \citep{Wood1987}. HM Cnc
has many weak helium emission lines, but with evidence for the
presence of a small amount of hydrogen \citep{Roelofs2010}. The
spectrum of V407 Vul appears similar to a G/K star spectrum and is
highly likely to be due to a chance alignment with a background star
\citep{Steeghs2006}. In \S 2 we noted that the optical colours for
\src\ are $g-r\sim$0.0 \citep{Haberl2017} and $u-g\sim$0.1--0.2
magnitudes.  These apparent colours are significantly less blue than
HM Cnc or ES Cet, or indeed most other AM CVn binaries. This maybe due
to a significant degree of reddening in the direction of \src: the
\citet{Schlegel1998} dust maps give a mean extinction to the edge of
the Galaxy of $E_{B-V}=0.45$. Even if the extinction to \src\ was half
this amount the dereddended colours would be similar to the known AM
CVn systems.

After their discovery there was much debate as to the nature of HM Cnc
and V407 Vul. Models included a white dwarf-white dwarf strongly
magnetic cataclysmic variable \citep{Cropper1998}; a white dwarf-red
  dwarf moderately magnetic cataclysmic variable \citep{Norton2004}; a
  unipolar inductor model in which electric currents heat up the more
  massive white dwarf so that X-rays are emitted \citep{Wu2002}; and a
  director impact model where an accretion stream impacts directly
  onto the more massive white dwarf without forming a disc
  \citep{Marsh2004}.  The mechanism for powering both sources remains
  an issue still to be resolved, although for HM Cnc the evidence
  seems to better support the direct impact model \citep{Roelofs2010}.

\begin{figure}
\begin{center}
\setlength{\unitlength}{1cm}
\begin{picture}(7,6.5)
\put(-1.5,-7.3){\includegraphics{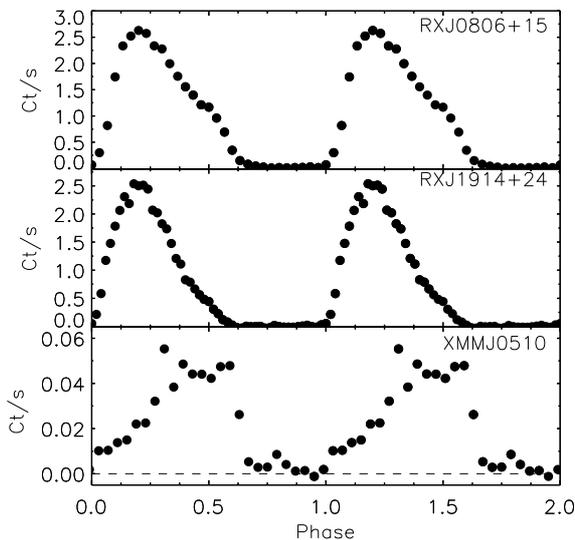}}
\end{picture}
\end{center}
\caption{The {\it XMM-Newton} (0.15-1keV) data of RX J0806+15 (top)
  folded on a period of 321 sec; RX J1914+24 (middle) folded on a
  period of 569 sec and \src\ (bottom) folded on a period of 1418
  sec.}
\label{xraycomp}
\end{figure}

\begin{figure*}
\begin{center}
\setlength{\unitlength}{1cm}
\hspace*{\fill}
\includegraphics[width=0.45\textwidth]{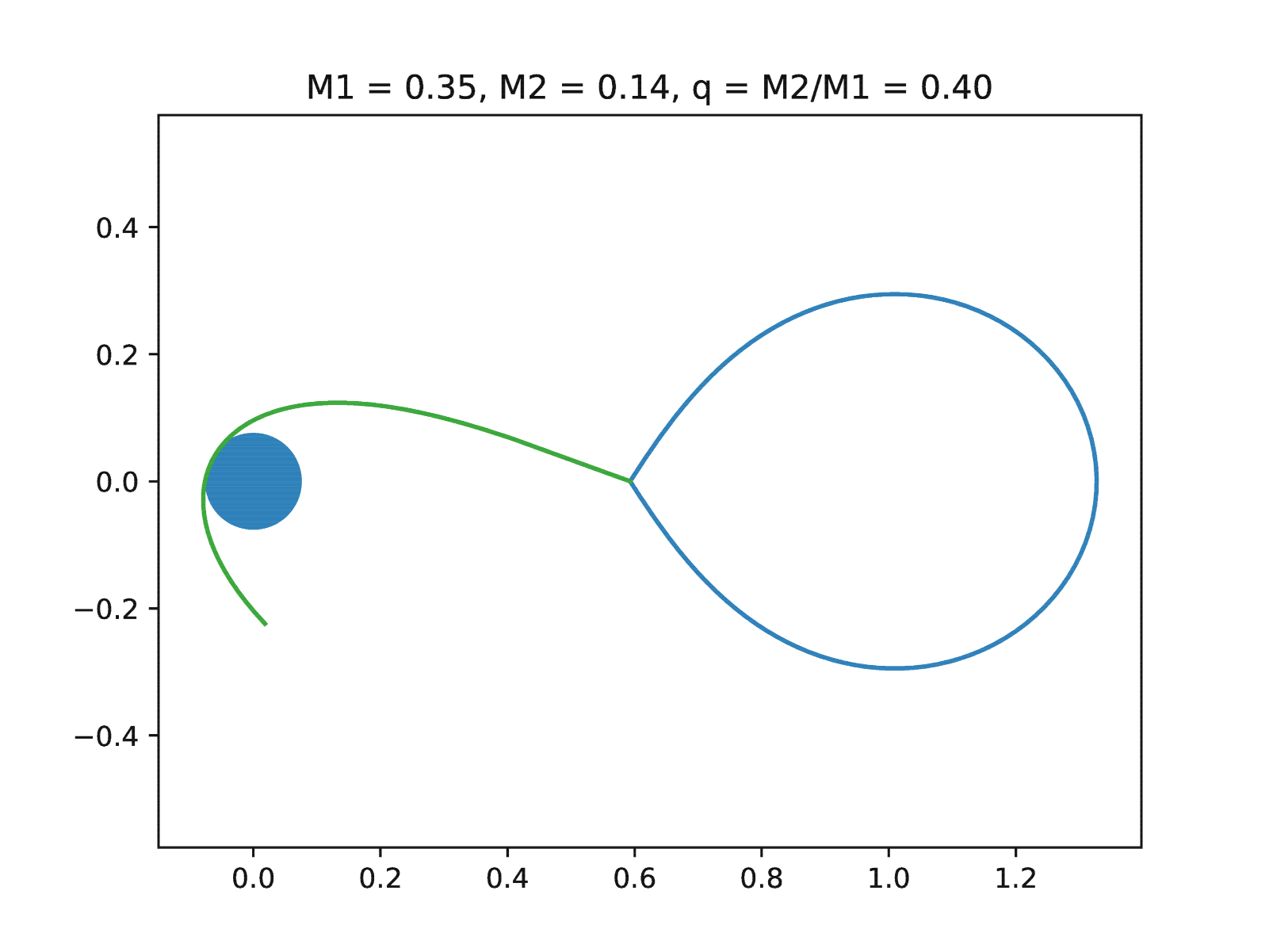}
\hspace*{\fill}
\includegraphics[width=0.45\textwidth]{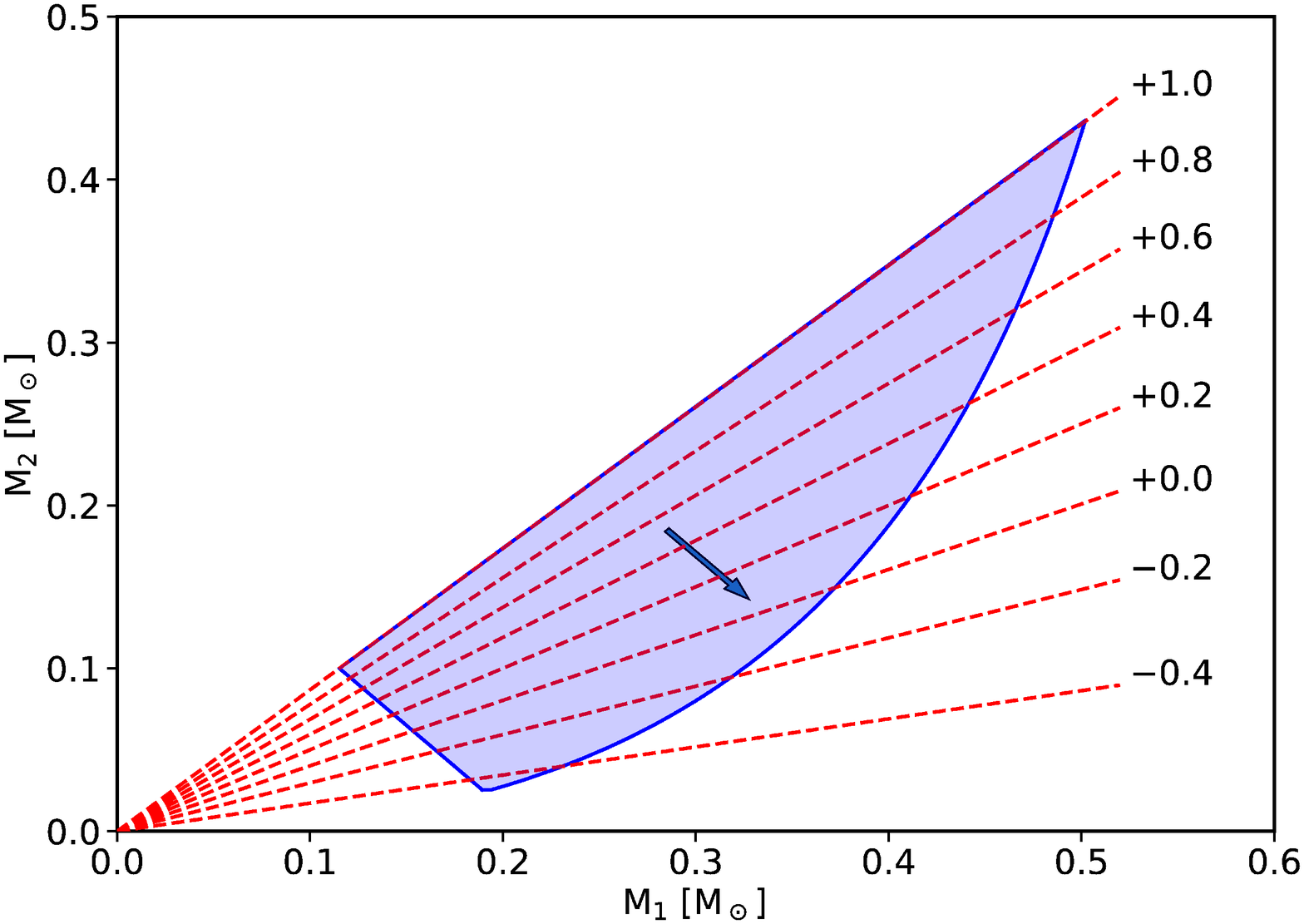}
\hspace*{\fill}
\end{center}
\caption{On the left-hand side we show an example of binary system
  with an orbital period of 23.6 min where direct impact accretion can
  occur: the parameters are $M_{1} =0.35\,$\Msun, $M_{2} =
  0.14\,$\Msun. The separation between the center of the two stellar
  components is defined as unity. On the right-hand side we show the
  region of $M_{1}$, $M_{2}$ values that lead to direct impact. Below
  the curved part of the boundary, the accretor is too small for a
  direct impact to occur (Eq.~31 from \citet{Marsh2004}. The diagonal
  sloping dashed lines mark the upper bound of stable accretion in the
  absence of tides for the different values of the logarithmic
  derivative of radius with respect to mass, $\zeta_2 =
  \mathrm{d}\ln(R_2)/\mathrm{d}\ln(M_2)$. We arbitrarily take the case
  $\zeta_2= 1$ to define the upper boundary which leads to $q < 0.87$.
  Models of helium star donors give lower values ($\zeta_2 \approx
  0.2$), while degenerate stars have $\zeta_2 < 0$, suggesting that
  the donor mass should lie in the lower half of the shaded
  region. The arrow shows the direction in which a given system would
  evolve in this diagram if undergoing conservative mass
  transfer. Along with the expectation that the donor must have
  started with mass $>0.1\,$\Msun\, this leads to the sloping boundary
  on the lower-left, although its exact location depends upon the
  assumed upper limit on $\zeta_2$.}
\label{directimpact}
\end{figure*}

To make a further comparison with HM Cnc and V407 Vul we show in
Figure \ref{xraycomp} the X-ray data in the 0.15--1keV band of HM Cnc,
V407 Vul and \src\ each folded on their X-ray periods (we extracted
these data from the {\sl XMM-Newton}
archive\footnote{http://nxsa.esac.esa.int} and reduced the data in the
same manner as \citet{Ramsay2005} and the most recent calibration
files). As noted by previous authors, the X-ray pulse profiles of HM
Cnc and V407 Vul are very similar, showing an on-off pulse where the
off phase shows essentially zero X-ray flux. This is similar to that
observed in \src. However, in contrast, compared to HM Cnc and V407
Vul, the rise to maximum in \src\ is more gradual and the decline is
more rapid -- almost the reverse of HM Cnc and V407 Vul. This could be
a clue to the location and extent of the emission region. For
instance, \citet{Dolence2008} and \citet{Wood2009} modelled the X-ray
and optical phase folded light curves to place constraints on the
extent of the X-ray generating spot assuming a direct impact model.

For a binary with an orbital period of 23.6 min, the mass accretor
requires a relatively large radius and hence low mass in order for the
accretion stream from the mass donor star to impact directly, although
this is ameliorated by the reduction in binary separation for fixed
period as the total system mass drops. To investigate this further we
took the direct impact model of \citet{Marsh2004} to place constraints
on the stellar masses. The left hand panel of
Figure~\ref{directimpact} shows an example of a binary with masses
($M_{1} =0.35\,$\Msun, $M_{2} = 0.14\,$\Msun) that can give rise to
direct impact on the more massive white dwarf. The right hand panel
shows the region of $M_1$, $M_2$ parameter space where, given some
assumptions about the nature of the donor star to be explained below,
stable direct impact accretion can occur. Not only must the system lie
in this region now, it should have started mass transfer inside it;
the direction of evolution assuming conservative mass transfer is
indicated by an arrow in Fig.~\ref{directimpact}. Assuming that the
donor must once have been at least $0.1\,$\Msun leads to the sloping
constraint on the left-hand side of the shaded regon. As expected, the
accretor has to be of low mass ($<0.5\,$\Msun) in order for direct
impact accretion to take place.

If the donor is completely degenerate, at a period of 23.6~min, it
would have a mass of $M_2 = 0.026\,$\Msun, a value achieved right at
the bottom tip of the shaded region with $M_1 \approx
0.20\,$\Msun. Although a degenerate donor is thus just about possible,
it seems likely that the donor has in fact significant entropy which
would avoid the requirement that it started mass transfer at a mass as
low as $0.1\,$\Msun.  This is in any case consistent with the finding
that much longer period AM~CVn stars contain donor stars of
significant entropy \citet{Green2018}. The response of the donor
star's radius to the loss of mass is important for the stability of
mass transfer. This is usually parameterised in terms of the
logarithmic derivative of radius with respect to mass $\zeta_2$ given
by
\begin{equation*}
\zeta_2 = \frac{\mathrm{d} \ln(R_2)}{\mathrm{d} \ln(M_2)}.
\end{equation*}
Degenerate stars expand upon loss of mass leading to $\zeta_2 < 0$
while non-degenerate stars have larger values of $\zeta_2$, which acts
to stabilise mass transfer. In the case of direct impact accretion,
loss of angular momentum from the orbit to the accretor is
destabilising, while tides transferring the angular momentum from the
accretor back to the orbit are stabilising \citep{Marsh2004}. In
Fig.~\ref{directimpact}, dashed lines parameterised by $\zeta_2$
indicate the upper limit for stable accretion in the absence of tides
(see \citet{Alexander1973}).  Low mass helium stars may make suitable
mass donors. \citet{Yungelson2008} presents models of such stars in
binaries which would fill their Roche lobe in a binary of period $P =
23.6\,$min when $M_2 \approx 0.06\,$\Msun\ and $R_2 \approx
0.05\,$\Rsun, at which point $\zeta_2 \approx
0.2$. Fig.~\ref{directimpact} shows that such a star could form the
donor in a stable direct impact accretion system if the accretor's
mass lay in the range $0.15$ to $0.30\,$\Msun. An alternative is that
that both the donor and accretor started as extremely low mass (ELM)
white dwarfs and the current system masses are $M_1 \approx
0.2\,$\Msun\ and $M_2 \approx 0.03\,$\Msun. The donor in this case
would be expected to have $\zeta_2$ close to the degenerate value,
$\approx -0.24$, which would transfer mass stably according to
Fig.~\ref{directimpact}. How the system is likely to evolve is a
separate question and beyond the scope of this paper. It should also
be remembered that tides could help stabilise systems that fail the
stability limits of Fig.~\ref{directimpact}.

The above discussion hinges on the direct impact interpretation of the
observational data. There are two important observational tests of the
direct impact model for \src. The first is the relative X-ray/optical
phase. In the direct impact model, the optical phase is determined by
the combination of light from the impact and from the
possibly-irradiated donor, while the X-ray flux should be dominated by
the impact region alone. We would always expect the optical flux to
peak prior to the X-ray flux, as seen in HM~Cnc and V407~Vul (Barros
et al 2007). However, comparing Fig.~\ref{directimpact} with Fig.~1 of
\citet{MarshSteeghs2002}, which examines the case for the shorter
period systems, indicates that \src's longer period may make the phase
difference relatively smaller than that seen in the shorter period
systems. The second test is to measure the rate of period change which
depends upon the component masses, $\zeta_2$ and the strength of any
tides. For instance, the expected period derivative switches from
positive to negative if $\zeta_2$ exceeds $1/3$. Given the long period
and faintness of \src, this will be a challenging but significant
constraint upon the nature of this interesting system. A third area of
interest will be searching for a periodic modulation in the optical
light curves at periods other than that of 23.6 min.

\section{Conclusions}

\src\ shows a periodic modulation in its X-ray flux on a period of
23.6 min. We find that the optical flux of \src\ is also modulated at
a period consistent with the X-ray period. The most likely explanation
is that this is an ultra-compact binary with a period of 23.6 min. The
X-ray folded light curve of \src\ shows similarities to that of HM Cnc
and V407 Vul although in \src\ the profile seems to be reversed
(i.e. there is a sharp decline towards minimum flux rather than from
minimum). \src\ and HM Cnc and V407 Vul all show a very soft X-ray
spectrum. Using only an assumption that 23.6 min is the orbital
period, we find that the direct impact accretor model can produce
stable accretion for a range of binary mass combinations. Low mass
helium stars or extremely low mass white dwarfs may be able to match
the properties needed of the donor in this system, but evolutionary
modelling is required to see if binaries of the right component masses
can be produced. The next goal observationally is to obtain
contemporaneous X-ray and optical data to allow a reliable phasing of
the multi-wavelength data to determine if they are offset as predicted
by the direct impact model. Given the faintness of the object it will
be a challenge to obtain phase-resolved optical spectroscopy to
identify radial velocity variations on a timescale of 23.6 min.

\section*{Acknowledgements}

This paper is based on observations collected at the European
Organisation for Astronomical Research in the Southern Hemisphere
under ESO programme 0102.D-0151. We thank Steven Parsons and Jay
Farihi for carrying out the observations. Some results presented in
this paper are based on observations obtained at the Gemini
Observatory, proposal ID GS-2016B-FT-25, which is operated by the
Association of Universities for Research in Astronomy, Inc., under a
cooperative agreement with the NSF on behalf of the Gemini
partnership: the National Science Foundation (United States), the
National Research Council (Canada), CONICYT (Chile), Ministerio de
Ciencia, Tecnolog\'{i}a e Innovaci\'{o}n Productiva (Argentina), and
Minist\'{e}rio da Ci\^{e}ncia, Tecnologia e Inova\c{c}\~{a}o
(Brazil). The authors thank the staff at the Gemini-South observatory
for performing the observations in service mode. Armagh Observatory
and Planetarium is core funded by the Northern Ireland Executive
through the Department for Communities.  TRM and DS thank STFC for
support via grants ST/L000733/1 and ST/P000495/1. VSD and ULTRACAM
thank STFC for support via consolidated grant ST/J001589.

\end{document}